\documentstyle[preprint,aps]{revtex}
\newcommand{\ra}{\rightarrow}
\newcommand{\be}{\begin{equation}}
\newcommand{\ee}{\end{equation}}
\newcommand{\om}{\omega}
\newcommand{\de}{\delta}
\newcommand{\ga}{\gamma}
\newcommand{\al}{\alpha}
\newcommand{\rS}{\stackrel{\ra}{S}}
\newcommand{\rr}{\stackrel{\ra}{r}}
\newcommand{\re}{\stackrel{\ra}{e}}
\newcommand{\rn}{\stackrel{\ra}{n}}
\newcommand{\vp}{\varphi}
\newcommand{\vth}{\vartheta}
\newcommand{\lal}{\langle\langle}
\newcommand{\rar}{\rangle\rangle}

\newcommand{\ve}{\varepsilon}
\newcommand{\rB}{\stackrel{\ra}{B}}

\begin{document}

\draft

\title{ ORIGIN OF PURE SPIN SUPERRADIANCE}
\author{V.I.Yukalov}
\address{Department of Mathematics and Statistics \\
Queen's University, Kingston, Ontario K7L 3N6, Canada \\
and \\
Bogolubov Laboratory of Theoretical Physics \\
Joint Institute for Nuclear Research, Dubna 141980, Russia}

\maketitle

\vspace{2cm}


\begin{abstract}

The question addressed in this paper is: What originates pure spin 
superradiance in a polarized spin system placed inside a resonator? The 
term "pure" means that no initial coherence is imposed on spins, and its 
appearance manifests a purely self--organized collective effect. The 
consideration is based on a microscopic model with dipole spin interactions. 
An accurate solution of evolution equations is given. The results show that 
the resonator Nyquist noise does not play, contrary to the common belief, 
any role in starting spin superradiance, but the emergence of the latter is 
initiated by local spin fluctuations. The decisive role of nonsecular dipole 
interactions is stressed.

\end{abstract}

A polarized spin system prepared in a nonequilibrium state returns to 
equilibrium through the spin--spin and spin--lattice relaxation mechanisms. 
The spin relaxation can be drastically accelerated if the nonequilibrium 
magnetic system is placed inside a coil of a resonance electric circuit with 
the natural frequency tuned to the precession frequency of spin magnetic 
moments [1]. The strong shortening of the relaxation time is caused by the 
coherence between individual rotating spins, which develops as a result of 
the interaction between the rotating magnetization and the resonator feedback 
field. This coherent phenomenon is analogous to the Dicke superradiance [2] 
occurring in atomic and molecular systems, and so Bloembergen and Pound [1] 
also called this fast collective damping in spin systems the radiation 
damping. Friedberg and Hartmann [3] noted that, in fact, the whole process 
in spin systems involves no radiation at all, but merely nonradiative transfer
of energy from the sample to the coil and back, the energy being dissipated 
in the circuit ohmically. And what one measures in experiments with spin 
systems is not the intensity of radiation, but the power of electric current. 
Nevertheless, the term superradiance has been accepted for the transient 
coherent phenomenon in spin systems, when it develops as a self--organized 
process, similarly to the Dicke superradiance. The similarity between the 
latter and the spin superradiance is not the sole excuse for the accepted 
term. Another reason is to distinguish the spin superradiance from nuclear 
induction, free or collective, that is also a coherent phenomenon although 
not self--organized, but forced by thrusting onto a sample an initial 
coherence. One more justification for using the term spin superradiance is 
that this effect is always accompanied by coherent magneto--dipole radiation, 
though the corresponding radiation intensity is very weak as compared to the 
easily measured current power [4].

The term {\it pure spin superradiance} is used in order to stress that this 
is a purely self--organized process, when coherence develops from an 
absolutely incoherent state. This is to be distinguished from {\it triggered 
spin superradiance} during which collective effects also play an important 
role, but the process starts from a coherent initial state, so that the 
imposed initial coherence triggers the development of a correlated state, in 
the same way as triggered optical superradiance [5] happens in atomic and 
molecular systems. When spin superradiance is caused by nuclear spins, it can 
be called the {\it nuclear spin superradiance}. A system of ion spins in a 
resonator cavity can, in principle, be also a source of spin superradiance.  

The  nuclear spin superradiance has been recently observed in a series of 
experiments [6-11] with different substances: from $\;Al\;$ nuclear spins in 
ruby ($\;Al_2O_3\;$) and from proton spins in propanediol ($\;C_3H_8O_2\;$), 
butanol ($\;C_4H_9OH\;$), and ammonia ($\;NH_3\;$). The interpretation of the 
pure spin superradiance in these experiments was based on the following 
picture. A system of polarized spins is placed in a constant magnetic field 
directed opposite to the sample polarization. This means that the spin system 
is prepared in an inverted state. The sample is put inside the coil of a 
passive electric circuit whose natural frequency is tuned to the Zeeman 
frequency of spins. Fluctuating magnetic field formed by the thermal Nyquist 
noise of the resonance circuit starts moving spins from their position of 
unstable equilibrium. The motion of spins induces electric current in the 
circuit, which creates a stronger magnetic field acting back on spins. Under 
the action of the feedback field, spins move faster increasing even more the 
resonator feedback field, and so on. This avalanche--type process results in 
a fast spin relaxation. Such is the commonly accepted picture of pure spin 
superradiance.

However in this, generally correct, picture there is one suspicious point, 
namely, that the beginning of the process is originated by the thermal Nyquist
noise of resonator. Really, if one attentively reads the classical paper by 
Bloembergen and Pound [1], then one finds there the estimate for the thermal 
damping, due to the thermal noise in resonator, showing that this damping is 
so negligibly small for macroscopic systems that it can never produce the 
initial thermal relaxation.

Thus we confront the alternative: either the common belief that this is the 
resonator thermal noise which initiates the pure spin superradiance is a 
delusion or Bloembergen and Pound are wrong. To resolve this paradox and to 
answer the question "what actually is the origin of pure spin superradiance" 
is the aim of the present paper.

The solution of the formulated problem meets the following difficulty. As 
follows from the analysis of Bloembergen and Pound [1], the homogeneous 
approach provided by the Bloch equation is not sufficient for correctly 
describing the process, but inhomogeneous local fields, that produce a 
microscopic relaxation mechanism, are essential. The phenomenological Bloch 
equation, even being solved in a reasonably accurate approximation [12], can 
describe only the triggered spin superradiance, when an initial coherence is 
imposed by assuming nonzero initial conditions for transverse magnetization. 
To take into account inhomogeneous local fields providing a microscopic 
relaxation mechanism means the necessity of dealing with a microscopic model. 
Writing the equations of motion for spin components we get a 
$\;3N\;$--dimensional system of nonlinear differential equations for $\;N\;$ 
spins, plus the Kirchhoff equation for an electric circuit. If one invokes 
any approach, to solve this system of equations, based on the uniform 
mean--field approximation, then one immediately returns to a homogeneous 
picture equivalent to the Bloch equation, thus loosing an information on 
local fields. When the number of spins, $\;N\;$, is not too large, say 
$\;N\sim 10-10^3\;$, then it is possible to resort to numerical calculations. 
Such a computer simulation has been done [4] (for mathematical details see 
[13]) and has been shown that, really, the pure spin superradiance can exist 
without the resonator Nyquist noise. However, such computer simulations have 
the following deficiencies: (i) they are time consuming; (ii) they are able 
to give only a qualitative description, since the number of spins involved 
is incomparably smaller than what one has in real samples with 
$\;N\sim 10^{23}\;$; (iii) they do not give analytical formulas that would 
be convenient to study with respect to the variation of all, sometimes 
numerous, parameters characterizing the system. Therefore, computer 
simulations can give a feel of what is happening, but cannot provide decisive 
answers. Below an analytical solution of microscopic equations is presented.

A system of nuclear spins is described [14] by the Hamiltonian
\be
\hat H =\frac{1}{2}\sum_{i\neq j}^{N}H_{ij} -\mu\sum_{i=1}^{N} \rB\rS_{i}
\ee
with dipole spin interactions
\be
H_{ij} =\frac{\mu^2}{r_{ij}}\left [ \rS_i\rS_j - 3\left (\rS_i\rn_{ij}
\right )
\left (\rS_j\rn_{ij}\right )\right ] ,
\ee
where $\;\mu\;$ is a nuclear magneton and
$$ \rn_{ij} \equiv \frac{\rr_{ij}}{r_{ij}}, \qquad 
\rr_{ij} \equiv \rr_i - \rr_j , \qquad r_{ij} \equiv |\rr_{ij}| . $$
The total magnetic field
\be
\rB = H_0\re_z + H\re_x
\ee 
consists of a constant external field $\;H_0\;$ and an alternating field 
$\;H\;$ of a resonator coil. The latter has $\;n\;$ turns of a cross--section 
area $\;A_c\;$ over a length $\;l\;$. The resonance electric circuit includes 
a resistance $\;R\;$, inductance $\;L\;$, and capacity $\;C\;$. The 
alternating resonator field
\be
H =\frac{4\pi n}{cl}j
\ee
is formed by an electric current satisfying the Kirchhoff equation
\be
L\frac{dj}{dt} +Rj +\frac{1}{C}\int_{0}^{t} j(\tau )d\tau = - 
\frac{d\Phi}{dt} + E_f ,
\ee
in which $\;E_f\;$ is an electromotive force and
$$ \Phi =\frac{4\pi}{c}nA_c\eta\rho \frac{\mu}{N} \sum_{i=1}^{N} 
\langle S_i^x \rangle $$
is a magnetic flux through the coil; $\;\eta \equiv V/V_c\;$ being a filling 
factor; $\;V_c \equiv lA_c\;$, the coil volume; $\;\rho \equiv N/V\;$, density
of spins.

Define the resonator natural frequency $\;\om \equiv 1/\sqrt{LC}\;$, ringing 
width $\;\ga_3\equiv R/2L\;$, and dimensionless fields
\be
h \equiv \frac{\mu H}{\hbar\ga_3} , \qquad 
f \equiv \frac{c\mu E_f}{nA_c\hbar\ga_3^2} .
\ee
Introduce the parameter
\be
\al_0 \equiv \pi\eta\frac{\rho\mu^2}{\hbar\ga_3}
\ee
characterizing the strength of coupling between the spin system and the 
resonator. Let us also use the notation
\be
u \equiv \frac{1}{N}\sum_{i=1}^{N}\langle S_i^-\rangle , \qquad
s \equiv \frac{1}{N}\sum_{i=1}^{N}\langle S_i^z\rangle
\ee 
for the mean spin components, where $\;\langle \ldots\rangle\;$ implies 
statistical averaging. Then the Kirchhoff equation (5) takes the form
\be
\frac{dh}{dt} +2\ga_3 h +\om^2\int_{0}^{t}h(\tau )d\tau = 
-2\al_0 \left ( u^* + u\right ) +\ga_3f .
\ee

To derive the evolution equation for the variables (8), we proceed as follows.
Write the Heisenberg equations for the corresponding spin components with the 
standard notation $\;\om_0\equiv\mu H_0/\hbar\;$ for the Zeeman frequency. 
Decouple the double spin correlators in the manner described by ter Haar 
[15], in order to preserve the terms containing the homogeneous spin--spin 
relaxation $\;\ga_2 \equiv T_2^{-1}\;$, which can be done by using 
second--order perturbation theory. For generality, we may also include the 
term describing the spin--lattice relaxation $\;\ga_1\equiv T_1^{-1}\;$. 
These steps are known and clear. The most difficult problem is how to treat 
the local spin fields
$$ \de_i =\frac{1}{\hbar}\sum_{j(\neq i)}^{N}\langle \frac{3}{2}a_{ij}S_j^z +
c_{ij}S_j^+ +c_{ij}^*S_j^-\rangle , $$
\be
\vp_i = -\frac{2}{\hbar}\sum_{j(\neq i)}^{N}\langle b_{ij}S_j^+ + 
c_{ij}S_j^z\rangle ,
\ee
caused by the dipole interactions
$$ a_{ij} =\frac{\mu^2}{r_{ij}^3}\left ( 1 - 3\cos^2\vth_{ij}\right ) , $$
\be
b_{ij} =-\frac{3\mu^2}{4r_{ij}^3}\sin^2\vth_{ij}
\exp \left (-i2\vp_{ij}\right ) ,
\ee
$$ c_{ij} =-\frac{3\mu^2}{4r_{ij}^3}\sin^2(2\vth_{ij})\exp \left ( 
-i\vp_{ij} \right ) , $$
where $\;\vth_{ij}\;$ and $\;\vp_{ij}\;$ are the spherical angles of 
$\;\rn_{ij}\;$. Note that in a uniform approximation the local fluctuating 
fields (10) are zero because of the properties of the dipole interactions 
(11).  To get a closed set of equations, at the same time retaining the 
information on the presence of fluctuating fields (10), we may replace the 
latter by stochastic fields, $\;\de_i\ra \vp_0, \; \vp_i \ra \vp\;$, the first
of which, in compliance with (10), is real and the second is complex. The 
distribution of these random fields is such that the averaging over it, which 
we shall denote by $\;\lal\ldots\rar\;$, gives
\be
\lal\vp_0\rar = \lal\vp\rar = 0 , \qquad \lal\vp_0^2\rar = 
\frac{1}{2}\lal|\vp |^2\rar =\ga_*^2 ,
\ee
where the dispersion $\;\ga_*\;$, in accordance with (10), is of the order of 
$\;\ga_2\;$. In this way, for the spin components (8) we obtain the system of 
stochastic equations
$$ \frac{du}{dt} = i(\om_0 -\vp_0 +i\ga_2)u - i(\ga_3h + \vp )s , $$
\be
\frac{ds}{dt} =\frac{i}{2} \left ( \ga_3h +\vp\right )u^* -
\frac{i}{2}\left (\ga_3 h + \vp^*\right ) u - \ga_1(s-\zeta ) ,
\ee
$$ \frac{d|u|^2}{dt}=-2\ga_2|u|^2 -i(\ga_3h +\vp )su^* +i(\ga_3h +
\vp^*)su . $$
The structure of (13) is transparent: $\;\ga_3h+\vp\;$ is the total effective 
field acting on spins; $\;h\;$ is the resonator field defined by (9); 
$\;\vp_0\;$ and $\;\vp\;$ model random local fields with a distribution whose 
particular form is not important since all we need is the property (12). If 
$\;\vp_0\;$ and $\;\vp\;$ were absent, then (13) would be reduced to the Bloch
equation.

To consider the case of pure spin superradiance, the initial conditions for 
the system of equations (9) and (13) are to be taken as
\be
h(0)=0 ,\qquad u(0)= 0, \qquad s(0)=z_0 .
\ee
The electromotive force $\;E_f = E_0\cos\om t\;$ in (5) corresponds to the 
resonance mode of the thermal Nyquist noise. The driving force in (6) is
\be
f=f_0\cos\om t ; \qquad f_0 \equiv \frac{c\mu E_0}{nA_c\hbar\ga_3^2} .
\ee

The system of equations (9) and (13) can be solved by a method [16] combining 
the guiding--center approach [17] and the method of averaging [18]. The idea 
is straightforward: First, we classify the variables onto fast and slow. To 
this end, we take into account the usual inequalities 
$\;\ga_1 \ll\om_0 ,\; \ga_2 \ll \om_0 ,\; \ga_3 \ll \om\;$, and consider 
the quasiresonance case, when $\;|\Delta | \ll \om_0\;$, where 
$\;\Delta \equiv \om -\om_0\;$ is detuning. Thence we notice right away 
that the variables $\;h\;$ and $\;u\;$ can be treated as fast, while $\;s\;$ 
and $\;|u|^2\;$ as slow. Keeping the latter as fixed parameters
\be
s=z , \qquad |u|=v ,
\ee
we get for the fast variables a system of linear equations, which, therefore, 
is not too difficult to solve. The found solutions for fast variables are to 
be substituted into the equations for slow variables, and the right--hand 
sides of these equations are to be averaged over the period $\;2\pi /\om_0\;$ 
of fast oscillations and also over the distribution of stochastic fields. This
procedure results in the equations
$$ \frac{dz}{dt} =g\ga_2w -\ga_1(z-\zeta ) -\ga_fz , $$
\be
\frac{dw}{dt} =-2\ga_2w -2g\ga_2wz +2\ga_fz^2
\ee
for the slow variables, where
$$ w \equiv v^2 -\frac{2\ga_*^2}{\om_0^2}z , \qquad 
g \equiv \al_0\left (\frac{\ga_3}{\ga_2}\right )
\frac{\pi (\ga_2 -\ga_3)^2}{(\ga_2 -\ga_3 )^2+\Delta^2} , $$
and the attenuation
$$ \ga_f \equiv \frac{f_0^2\ga_3^4}{32\om_0^2(\Delta^2+\ga_2^2)}
\left \{ \left ( 1 +\frac{8\pi^2}{3}\right )\ga_2 -2\pi\Delta +\right. $$
\be
\left. +\frac{\om_0z}{\Delta^2 +\ga_2^2}\left [ (\al -2\pi\beta )(\Delta^2 -
\ga_2^2) +2\ga_2\Delta (\beta +2\pi\al )\right ]\right \} ,
\ee
in which 
$$ \al \equiv \al_0 \left (\frac{\ga_3}{\om_0}\right ) 
\frac{\pi (\ga_2 -\ga_3)^2}{(\ga_2 -\ga_3)^2+\Delta^2} , \qquad
\beta \equiv \al_0 \left (\frac{\ga_3}{\om_0}\right ) 
\frac{\pi (\ga_2 -\ga_3)\Delta}{(\ga_2 -\ga_3)^2+\Delta^2} , $$
is due to the action of the driving field (15).

The amplitude of the electromotive force related to the thermal Nyquist noise 
[19], at temperature $\;T\;$ satisfying the inequality 
$\;\hbar\om \ll k_BT\;$, is given by $\;E_0^2 =\ga_3Rk_BT/\pi\;$. Whence 
for the amplitude of the driving field (15) we have
\be
f_0^2 =\frac{8\al_0k_BT}{\pi\hbar\ga_3N} .
\ee
Let us accept the values of parameters characteristic of experiments [7-11] 
with proton spins: $\;\om_0 \sim \om \sim 10^8sec^{-1},\; 
\ga_1 \sim 10^{-5}sec^{-1}, \; \ga_2 \sim 10^5sec^{-1},\; 
\ga_3 \sim 10^6sec^{-1}, \; T \sim 0.1K\;$, and $\;N \sim 10^{23}\;$. Then 
$\;f_0\sim 10^{-10}\;$ and the thermal attenuation (18) is 
$\;\ga_f \sim 10^{-16}sec^{-1}\;$. Such an insignificant quantity, of course, 
plays no role, as compared to all other damping parameters, and has to be 
neglected in (17).

This result shows, in agreement with Bloembergen and Pound [1], that the 
Nyquist noise of resonator can never produce the initial thermal relaxation, 
thus, cannot originate the pure spin superradiance.

Ommitting in (17) the negligibly small $\;\ga_f\;$ and taking into account 
that $\;\ga_1 \ll \ga_2\;$, we come to
\be
\frac{dz}{dt} =g\ga_2w , \qquad \frac{dw}{dt} =-2\ga_2w(1 +gz) .
\ee
According to (14), the initial conditions are $\;z(0)=z_0\;$ and 
$\;v(0)=0\;$. Equations in (20) are exactly integrable yielding
$$ z=\frac{\ga_0}{g\ga_2}{\rm tanh}\left (\frac{t-t_0}{\tau_0}\right ) -
\frac{1}{g} , $$
\be
v^2 =\left (\frac{\ga_0}{g\ga_2}\right )^2{\rm sech}^2\left (
\frac{t-t_0}{\tau_0}\right ) +\frac{2\ga_*^2}{\om_0^2}z ;
\ee
here $\;\ga_0\;$ is the radiation width given by
\be
\ga_0^2 =\Gamma_0^2 -2(g\ga_2)^2\ve_*z_0 ,
\ee
where
$$ \Gamma_0 \equiv \ga_2(1+gz_0) , \qquad \ve_* \equiv 
\left (\frac{\ga_*}{\om_0}\right )^2 ; $$
the radiation time $\;\tau_0 =\ga_0^{-1}\;$; and the delay time is 
\be
t_0 =\frac{\tau_0}{2}\ln 
\left |\frac{\ga_0 -\Gamma_0}{\ga_0 +\Gamma_0}\right | .
\ee

The criterion for the occurrence of spin superradiance is the validity of the 
inequalities
\be
0 < t_0 < \infty , \qquad \tau_0 < T_2 .
\ee
Invoking (22) and (23) and bearing in mind that $\;\ve_* \ll 1\;$, we find 
that (24) is equivalent to
\be
z_0 < z_p \equiv  -\frac{2}{g}, \qquad \ve_* > 0.
\ee
As far as $\;|z_0| < 1/2\;$, the first of the inequalities (25) requires that
$\;g \geq 4\;$. In this way, the pure spin superradiance occurs when the 
initial spin polarization $\;z_0\;$ is negative, with an absolute value 
surpassing the threshold $\;|z_p|=2/g\;$, when the coupling of the spin 
system with a resonator is sufficiently strong, $\; g\geq 4\;$, and if there 
exist local random fields with a nonzero dispersion $\;\ga_* > 0\;$.

To emphasize the crucial importance of the local fields, let us notice that 
if one puts $\;\ve_* \ra 0\;$, then $\;\ga_0 \ra |\Gamma_0|\;$ and 
$\;|t_0|\ra \infty\;$. That is, without these fields the pure spin 
superradiance is impossible. To make the essential dependence of the delay 
time on $\;\ve_*\;$ apparent, we may write (23) for the case of strong 
coupling, when $\;g|z_0|\gg 1\;$, then
$$ t_0 \simeq \frac{T_2}{2g|z_0|}\ln\left | \frac{2z_0}{\ve_*}\right | . $$
From here it is evident that $\;t_0 \ra \infty\;$ as $\;\ve_* \ra 0\;$. For 
the parameters typical of experiments [7-11] we have 
$\;t_0 \sim 10^{-6}-10^{-5}sec\;$. So, this is the local random fields 
that are responsible for starting the process of self--organization leading 
to the pure spin superradiance.

One more question is worth answering: Which part of the local fields is more 
important for initiating the pure spin superradiance? Recall that the 
stochastic fields entering into the evolution equations (13) are related to 
two types of local fields defined in (10). As follows from (13), the term 
$\;\de_i\;$ in (10), corresponding to $\;\vp_0\;$, only shifts the rotation 
frequency, while the term $\;\vp_i\;$, corresponding to $\;\vp\;$, starts 
moving the spin $\;z\;$--component even when the resonator feed back field 
$\;h\;$ is yet absent. The term $\;\vp_i\;$ in (10) is due to the dipole 
interactions $\;b_{ij}\;$ and $\;c_{ij}\;$ defined in (11). These 
interactions, in the theory of magnetic resonance [14], are called nonsecular 
interactions, as compared to the secular interaction $\;a_{ij}\;$. The initial
motion of spins, when $\;u(0)=0\;$ and $\;h(0)=0\;$, is due solely to the 
action of nonsecular interactions. This conclusion is in agreement with 
computer simulations [4,20] for small and mesoscopic spin systems with 
$\;N\sim 10-10^3\;$. Thus, we are in a position to give the final answer to 
the question posed in this paper:

The pure spin superradiance in a nonequilibrium system of polarized nuclear 
spins can be originated only by local fields due to nonsecular dipole 
interactions. The thermal Nyquist noise of resonator plays no role in this 
process.

\vspace{5mm}

I would like to express my sincere gratitude to R.Friedberg, S.R.Hartmann, 
and J.T.Manassah for useful discussions and helpful advises, as well as for 
their kind hospitality during my visits to the Columbia University and City 
University of New York. Financial support from the Natural Sciences and 
Engineering Research Council of Canada is appreciated.

\end{document}